\documentclass[11 pt]{amsart}
\usepackage{amsmath}
\usepackage{amssymb}
\usepackage{amsthm}
\usepackage{dsfont}
\usepackage{fancyhdr}
\usepackage[all]{xy}
\usepackage{hyperref}
\usepackage{paralist}
\usepackage{tikz}
\usepackage{verbatim}
\usepackage[hmargin=1.1in,vmargin=1.1in]{geometry}

\newtheorem{theorem}{Theorem}[section]

\theoremstyle{proposition}

\theoremstyle{corollary}

\theoremstyle{definition}

\numberwithin{equation}{section}

\newcommand{\Z}{\mathbb Z}

\newcommand{\cL}{{\mathcal L}}
\newcommand{\F}{\mathbb F}

\newcommand{\Q}{\mathbb  Q}

\newcommand{\End}{\rm End}
\newcommand{\Pic}{\rm Pic}

\newcommand{\Div}{\rm div}
\newcommand{\nuinf}{\nu_{\infty}}
\newcommand{\siminf}{\sim_{\infty}}

\begin{document}
\title[]{Trilinear maps for cryptography II}
\author[]{Ming-Deh A. Huang (USC, mdhuang@usc.edu)}
\address{Computer Science Department,University of Southern California, U.S.A.}
\email{mdhuang@usc.edu}

\urladdr{}

\maketitle

\begin{abstract}
We continue to study the construction of cryptographic trilinear maps involving abelian varieties over finite fields.  We introduce Weil descent as a tool to strengthen the security of a trilinear map.
We form the trilinear map on the descent variety of an abelian variety of small dimension defined over a finite field of a large extension degree over a ground field.   The descent bases, with respect to which the descents are performed, are  trapdoor secrets for efficient construction of the trilinear map which pairs three trapdoor DDH-groups.  The trilinear map also provides efficient public identity testing for the third group.
We present a concrete construction involving the jacobian varieties of hyperelliptic curves.
\end{abstract}
\section{Introduction}
Cryptographic applications of multilinear maps were first proposed in the work of  Boneh and Silverberg \cite{BS}.  However the existence of cryptographically interesting $n$-linear maps for $n > 2$ remains an open problem. The problem has attracted much attention more recently as multilinear maps and their variations have become a useful tool for indistinguishability obfuscation. Very recently Lin and Tessaro \cite{LT} showed that trilinear maps are sufficient for the purpose of achieving indistinguishability obfuscation (see \cite{LT} for references to related works along several lines of investigation).

In this paper we continue to study cryptographic trilinear maps involving abelian varieties over finite fields along the line of investigation started in \cite{H}. This line of investigation was motivated by an observation of Chinburg (at the AIM workshop on cryptographic multilinear maps (2017)) that the following map from \'{e}tale cohomology may serve as the basis of constructing a cryptographically interesting trilinear map:

\[ H^1 (A,\mu_{\ell})\times H^1 (A,\mu_{\ell})\times H^2 (A,\mu_{\ell})\to H^4 (A, \mu_{\ell}^\otimes{3})\cong \mu_{\ell}\]
where $A$ is an abelian surface over a finite field $\F$ and the prime $\ell\neq {\rm char}(\F)$.  This trilinear map is the starting point of the following more concrete construction.

Suppose $A$ is a principally polarized abelian variety over a finite field $\F$. Let $A^*$ denote the dual abelian variety. Consider $A$ as a variety over $\bar{\F}$, the algebraic closure of $\F$.
Let $e_{\ell}$ be the pairing between $A[\ell]$ and $A^* [\ell]$ (\cite{MilneA} \S~16).

In \cite{H} we consider the trilinear map $(\alpha,\beta,\cL) \to e_{\ell}(\alpha,\varphi_{\cL} (\beta))$, where  $\alpha,\beta\in A[\ell]$, $\cL$ is an invertible sheaf, and $\varphi_{\cL}$ be the map $A\to A^* =\Pic^0 (A)$ so that
\[ \varphi_{\cL} (a) =t_a^*\cL \otimes \cL^{-1} \in \Pic^0 (A)\]
for $a\in A(\bar{\F})$ where $t_a$ is the translation map defined by by $a$ (\cite{MilneA} \S~1 and \S~6).

Note that in the map just described we no longer need to assume that $A$ is of dimension 2.

\section{General construction}
We describe the general idea of constructing a cryptographic trilinear map motivated by the above discussion.

We assume that $A$ is a simple and  principally polarized abelian variety defined over a finite field.
Let $e: A[\ell]\times A[\ell]\to\mu_{\ell}$ be a non-degenerate skew-symmetric pairing.  An important example is the pairing defined by a polarization of $A$ and the canonical pairing between $\ell$-power torsion points of $A$ and the dual abelian variety.  We refer to \cite{H} for a description in the context of constructing trilinear maps.

To construct a trilinear map we find $\alpha,\beta\in A[\ell]$ such that
$e(\alpha,\beta) \neq 1$.

We form a nontrivial submodule $U$ of the module
$W=\{ \lambda\in\End (A[\ell]):
e (\alpha,\lambda(\beta) )=1\}$, and let $U_1$ be the module generated by 1 and elements of $U$.

Let $G_1$ and $G_2$ be respectively the cyclic groups generated by $\alpha$ and $\beta$, and
 $G_3 = U_1/U$ with $1+U$ as the generator, we consider the trilinear map $G_1\times G_2 \times G_3 \to \mu_{\ell}$ sending
$(x\alpha,y\beta,z+U)$ to $\zeta^{xyz}$ where $\zeta=e(\alpha,\beta)$.  The map is well defined since for $\lambda\in U$, $e(\alpha,\lambda(\beta))=1$.

For the trilinear map to be cryptographically interesting we need the map to be efficiently computable on the one hand, and the discrete logarithm problems on $G_i$, $i=1,2,3$, should be hard to solve on the other hand.  For the trilinear map to be efficiently computable, we need the pairing $e$ to be efficiently computable, we also need a representative of $z+U$ to be efficiently specified and efficiently executable when applied to the group $G_2$.

The pairing computation in general takes time at least exponential in the dimension of $A$.
Therefore in our earlier construction in \cite{H}, we assume that the dimension of $A$ small.  In \cite{H}, $U_1$ is formed from the endomorphism ring of $A$. In this situation the discrete logarithm problem on $G_3$ is potentially vulnerable to a line of attack using trace pairing.  The trace pairing can in principle be reduced to intersection product, assuming the polarization divisor can be efficiently presented.  These computations may be considered efficient when $\dim A$ is fixed, even though they can be exponential in $(\dim A)^2$, or even worse.

To strengthen the trilinear map we apply Weil descent \cite{A,FL} to effectively raise a barrier of dimension. We form the trilinear map on the descent variety $\hat{A}$ of an abelian variety $A$ of small dimension.  If $\dim A = g$ and $A$ is defined over a field $K$ of extension degree $d$ over $k$, then $\hat{A}$ is defined over $k$ of dimension $dg$.  The descent bases, with respect to which the descents are performed, are  trapdoor secrets for efficient construction of the trilinear map which pairs three trapdoor DDH-groups.  Different descent bases are used for the first group and the second group, so as to prevent the DDH problem from an attack that utilizes the published pairing to induce self-pairing.  The trilinear map also provides efficient public identity testing for the third group.

Frey \cite{F} introduced Weil descent as a constructive tool in cryptography to disguise elliptic curves.  In \cite{DG}
Dent and Galbraith applied the idea to construct hidden pairings based on which trapdoor DDH groups can be constructed.
The construction in \cite{DG} that involves Weil descent is vulnerable to the attacks described in \cite{Mo}.  Those attacks depend critically on the addition map of interest being given in the projective model by homogeneous polynomials.
The attacks do not extend to constructions involving Weil descent where the abelian varieties and maps are given strictly by affine models in affine pieces, such as ours.
In our framework,  a construction similar to that in \cite{DG} but more restrictive would be to
select a secret descent basis of a finite field $K$ of a large extension degree over a smaller finite field $k$, and
an elliptic curve $E$ defined over $K$, then make public the descent of a $K$-rational point of $E$, together with the descent maps for addition morphism and doubling morphism.  In this case it follows directly from Theorem~\ref{gld} in \S~\ref{secrecy} that the descent basis can be uncovered from the published maps, hence the scheme is not secure.  The analysis in \S~\ref{analysis} also provides other reasons why in our framework constructions based on elliptic curves are not secure.

After a brief discussion of Weil descent in the next section we will proceed to the trilinear map construction involving Weil descent in \S~\ref{tri-weil}.  We study the security issues in specifying descent maps in \S~\ref{secrecy}, and in \S~\ref{jac} we present a concrete trilinear map construction using the jacobian varieties of hyperelliptic curves.

\section{Weil descent}
\label{weil}
Let $k$ be a finite field and let $K$ be an extension of degree $d$ over $k$.
Let $u_1,\ldots,u_d$ be a basis of $K$ over $k$.  Then
\[ u_i u_j = \sum_{i=1}^d \delta_{ijk} u_k\]
with $\delta_{ijk}\in k$ for $1\le i,j,k\le d$.
The indexed set $\delta_{ijk}$, denoted $\Delta$, is called the descent table with respect to the basis $u_1,\ldots,u_d$.
More generally, for $k\ge 2$,
\[ u_{i_1}\ldots u_{i_k} = \sum_{j=1}^d \delta_{i_1,\ldots,i_k, j} u_j\]
with $\delta_{i_1,\ldots,i_k,j}\in k$, and $1\le i_1,\ldots,i_k \le d$.
The indexed set $\delta_{i_1,\ldots,i_k,j}$, denoted $\Delta^{(k)}$, is called the $k$-th descent table with respect to the basis $u_1,\ldots,u_d$. The table $\Delta^{(k)}$ can be easily derived from the $\Delta$ and $\Delta^{(k-1)}$.
For example $\delta_{ijks} = \sum_r \delta_{ijr}\delta{rks}$

Let $R=K[x_1,\ldots,x_n]$. Suppose $F\in R$. Consider the substitution of variables
$x_i = \sum_{j=1}^d y_{ij} u_j$, for $i=1,\ldots,d$.
Let $\hat{R}=k[y_{11},\ldots,y_{nd}]$.  Denote by $\tilde{F}$ the polynomial in $\hat{R}$ obtained from $F$ by substituting $x_i$ with $\sum_{j} y_{ij} u_j$.  Thus
\[ \tilde{F} (\hat{x}_1,\ldots,\hat{x}_n): = F (\tilde{x}_1,\ldots,\tilde{x}_n)=\sum_{i=1}^d f_i u_i \]
where $\hat{x}_i= (y_{ij})_{j=1}^d$, $\tilde{x}_i= \sum_{j=1}^d y_{ij}u_j$, and $f_i (\hat{x}_1,\ldots,\hat{x}_n)\in k[\hat{x}_1,\ldots,\hat{x}_n]=k[y_{11},\ldots,y_{nd}]$.

We call $(f_i)_{i=1}^d$ the {\em descent} of $F$ with respect to the basis $u_1,\ldots,u_d$, denoted as $\hat{F}$.
They are easy to construct with the help of the descent table.

We also refer to $\hat{F}$ as the {\em global descent} of the polynomial $F$.

Write $F=\sum_i t_i$ where $t_i$ is a term, that is a constant times a monomial.  Then $\hat{F}$ contains the descent of
$\hat{t_i}$ for all $i$. In terms of vector summation we may write $\hat{F}=\sum_i \hat{t_i}$.

For this section we fix a basis $u_1,\ldots,u_d$ and we will omit the phrase "with respect to the basis $u_1,\ldots,u_d$" when defining descent objects.

Let $\delta$ denote the linear map $\bar{k}^d \to \bar{k}$ such that if we put
$x=(x_i)_{i=1}^d$,
$\delta(x) = \sum_{i=1}^d x_i u_i$.

Let $\sigma$ be a generator of $G(K/k)$ (for example the Frobenius automorphism over $k$), the Galois group of $K/k$.
Then $\delta^{\sigma^j} (x) = \sum_{i=1}^d x_i u_i^{\sigma^j}$ for $j=0,\ldots,d-1$.

Let $\rho$ denote the bijective linear map
$\bar{k}^d \to \bar{k}^d$ such that $\rho (x) = (\delta^{\sigma^i} (x) )_{i=0}^{d-1}$ for $x=(x_i)_{i=1}^d\in \bar{k}^d$.

Let $\hat{x}_i = (x_{ij})_{j=1}^d$, for $i=1,\ldots,n$.
Then
\[ F (\delta (\hat{x}_1),\ldots,\delta (\hat{x}_n) ) = \sum_{i=1}^d f_i (\hat{x}_1,\ldots,\hat{x}_n) u_i .\]
In fact, for $j=0,\ldots, d-1$,
\[ F^{\sigma^j} (\delta^{\sigma^j} (\hat{x}_1),\ldots,\delta^{\sigma^j} (\hat{x}_n) ) = \sum_{i=1}^d f_i (\hat{x}_1,\ldots,\hat{x}_n) u_i^{\sigma^j} .\]

If we identify $\bar{k}^{dn}$ as the $n$-fold product $\bar{k}^d \times \ldots \times \bar{k}^d$, and by abuse of notation denote $\delta$ as the map $\bar{k}^{dn} \to \bar{k}^n$ such that
$\delta (\beta_1,\ldots,\beta_n) = (\delta (\beta_1),\ldots, \delta(\beta_n))$ where $\beta_1,\ldots,\beta_n\in\bar{k}^d$.
Then we may write
$F\circ\delta =\delta\circ\hat{F}$.  In fact,
$F^{\sigma^j}\circ\delta^{\sigma^j} =\delta^{\sigma^j}\circ\hat{F}$ for $j=0,\ldots,d-1$.

Let $\iota: \bar{k}\to\bar{k}^d$ be such that $\iota(\alpha)= (\sigma^i (\alpha))_{i=0}^{d-1}$ for $\alpha\in\bar{k}$.
For $\alpha\in\bar{k}$, we define the {\em descent} of $\alpha$ to be the
unique $\beta \in \bar{k}^d$, such that
$\rho (\beta) = \iota (\alpha)$.
We have $\sigma^j (\alpha) = \delta^{\sigma^j} (\beta)$ for $j=0,\ldots,d-1$. In particular $\alpha = \delta (\beta)$.
If $\alpha\in K$, then $\alpha=\sum_{i=1}^d a_i u_i$ with $a_i \in k$.  In this case $\hat{\alpha} = (a_i)_{i=1}^d$.
Hence there is a bijection between $K\to k^d$ sending $\alpha\in K$ to $\hat{\alpha}$.

More generally if $\alpha=(\alpha_i)_{i=1}^n \in \bar{k}^n$, then its descent, $\hat{\alpha}$, is $(\hat{\alpha_i})_{i=1}^n$, which we consider an element in $\bar{k}^{nd}$.

If $V = Z(F)$, the algebraic set defined by the zeroes of some $F\in R$, then its {\em descent} is $\hat{V}:=Z(f_1,\ldots,f_d)$ where $\hat{F}=(f_i)_{i=1}^d$.  From the discussion above we see that if $\alpha\in\hat{V} (\bar{k})$ then
$F^{\sigma^j} (\delta^{\sigma^j} (\alpha) ) = 0$, so $\delta^{\sigma^j} (\alpha) \in V^{\sigma^j} (\bar{k})$.
This shows that $\rho ( \hat{V}(\bar{k}) )\subset \prod_{i=0}^{d-1} V^{\sigma^j} (\bar{k})$.
Conversely if $\alpha=(\alpha)_{i=0}^{d-1}\in \prod_{i=0}^{d-1} V^{\sigma^j} (\bar{k})$, let $\beta\in\bar{k}^d$ such that
$\rho (\beta) = \alpha$.  Then for $i=0,\ldots,d-1$, $\delta^{\sigma^i} (\beta)=\alpha_i\in V^{\sigma^i}(\bar{k})$, so $F^{\sigma^i}(\alpha_i ) =0$, so
$\delta^{\sigma^i} (\hat{F} (\beta)) = F^{\sigma^i} (\delta^{\sigma^i} (\beta)) = 0$.   We have $\rho (\hat{F} (\beta)) =0$, so $\hat{F} (\beta) = 0$,  so $\beta\in \hat{V}(\bar{k})$.
It follows that
$\rho$ restricts to a linear bijection between $\hat{V}(\bar{k})$ and $\prod_{i=1}^{d-1} V^{\sigma^j} (\bar{k})$.

For $\alpha\in K^n$, $\widehat{F(\alpha)} = \hat{F} (\hat{\alpha})$, it follows that
there is a bijection between $V(K)$ and $\hat{V}(k)$ sending a $K$-point in $V$ to its descent, which is in $\hat{V}(k)$.

More generally suppose $V=Z(F_1,\ldots,F_m)$, the algebraic set defined by the zeroes of $F_1,\ldots,F_m \in R$.  Suppose $\hat{F_i} = (f_{ij})_{j=1}^d$ for $i=1,\ldots,d$.  Then the {\em descent} of $V$ is $\hat{V}=Z(f_{11},\ldots,f_{md})$.
Similarly $\rho$ restricts to a linear bijection between $\hat{V}(\bar{k})$ and $\prod_{i=1}^{d-1} V^{\sigma^j} (\bar{k})$, and there is a bijection from $V(K)$ to $\hat{V}(k)$.

Let the natural extension of $\iota$ to $\bar{k}^n \to \bar{k}^{nd}$ be denoted by $\iota$ as well.  Consider the restrictions of maps $ \delta: \hat{V} \to V$, $\iota: V \to \prod_{i=1}^{d-1} V^{\sigma^i}$ and
$\rho: \hat{V} \to \prod_{i=1}^{d-1} V^{\sigma^i}$.  We have $\iota \circ \delta =\rho$.  For $\alpha\in V(\bar{k})$, $\hat{\alpha}$ is the unique $\beta\in \hat{V} (\bar{k})$ such that $\rho (\beta) = \alpha$.  In particular,
$\delta(\beta) = \alpha$.

Suppose $\varphi$ is an algebraic map from $V(\bar{k})$ to $\bar{k}$ defined over $K$.
The {\em descent} of $\varphi$, denoted $\hat{\varphi}$, is the map $\hat{V} (\bar{k}) \to \bar{k}^d$ defined over $k$ such that
$\delta\circ\hat{\varphi} = \varphi\circ\delta$.
Since $\hat{\varphi}$ is defined over $k$, it follows that $\delta^{\sigma^i}\circ\hat{\varphi} = \varphi^{\sigma^i}\circ\delta^{\sigma^i}$ for $i=0,\ldots,d-1$.
We have $\rho\circ \hat{\varphi} = (\prod_{i=0}^{d-1} \varphi^{\sigma^i}) \circ \rho$.  This also justifies the uniqueness of $\hat{\varphi}$.

So
for $(\beta_1,\ldots,\beta_n)\in \hat{V}(\bar{k})$ with $\beta_1,\ldots,\beta_n \in \bar{k}^d$,
\[ \delta^{\sigma^i} ( \hat{\varphi} (\beta_1,\ldots,\beta_n)) = \varphi^{\sigma^i} (\delta^{\sigma^i}( \beta_1),\ldots,\delta^{\sigma^i}(\beta_n) ).\]

The {\em descent function} of $\varphi$, denoted $\tilde{\varphi}$, is the map (function) $\delta\circ \hat{\varphi} :\hat{V} (\bar{k}) \to \bar{k}$.

More generally if $\varphi$ is a map $V (\bar{k})\to \bar{k}^r$ with $\varphi_i$ as the $i$-th coordinate map so that
$\varphi (\alpha) = (\varphi_ (\alpha))_{i=1}^r$.  The {\em descent} of $\varphi$, denoted $\hat{\varphi}$, is the map $\hat{V} (\bar{k}) \to \bar{k}^{rd}=\bar{k}^d\times\ldots\times\bar{k}^d$ such that
$\delta \hat{\varphi} = \varphi\delta$.
We have $\rho\circ \hat{\varphi} = (\prod_{i=0}^{d-1} \varphi^{\sigma^i}) \circ \rho$.

\section{Trilinear maps involving Weil descent}
\label{tri-weil}
\subsection{Constructing the trilinear map}
 Starting with an abelian  variety $A$ of dimension $g$ defined over a finite field $K$ of extension degree $d$ over a finite field $k$,
we consider the descent $\hat{A}$ of $A$ with respect to a basis $u_1,\ldots,u_d$ of $K$ over $k$.

For simplicity assume  $\log\ell$, $d$ and $\log | k |$ are linear in the security parameter $n$, whereas $g=O (\log^{1-\epsilon} n)$. We use the descent basis to construct and specify a trilinear map efficiently.    However the abelian variety $A$ as well as the descent basis will be kept secret.

 Let $\delta$ and $\rho$ be the maps defined in \S~\ref{weil}, which are determined by the descent basis.   So $\rho$ induces an isomorphism $\rho:\hat{A}\to \prod_{i=0}^{d-1} A^{\sigma^i}$ defined over $K$ where $\sigma$ is a generator of the Galois group of $K$ over $k$.

We have $\hat{A}[\ell] \stackrel{\rho}{\cong} \prod_{i} A^{\sigma^i}[\ell]$ and
$\prod_i A^{\sigma^i}[\ell] \cong   A[\ell]^d$.
With the identification of $\hat{A}[\ell]$ and $A[\ell]^d$,
a $d\times d$ matrix $M=(a_{ij})$ with $a_{ij}\in\F_{\ell}$ defines an element $\varphi_M \in \End (\hat{A}[\ell])$, so that
if $D\in \hat{A}[\ell]$ is identified with $(D_i)_{i=0}^{d-1}$  with $D_i \in A[\ell]$, then
$\varphi_M (D)$ is identified with $M (D_i)_{i=0}^{d-1}$.

\[
\begin{array}{llll}
\hat{A}[\ell] & \stackrel{\rho}{\to} & \prod_{i} A^{\sigma^i} [\ell]\cong  & A[\ell]^d\\
\downarrow \varphi_M & &   & \downarrow M\\
\hat{A}[\ell] & \stackrel{\rho}{\to} & \prod_{i} A^{\sigma^i} [\ell]\cong  & A[\ell]^d

\end{array}
\]

To construct a trilinear map,
we form a a set of $N_1=d^{O(1)}$ maps $\varphi_i$ where $\varphi_i = \varphi_{M_i}$ for some $d\times d$, (0,1)-matrix such that
(1) there are exactly two nonzero entries $(i,i_1)$ and $(i,i_2)$ for row $i$, for $i=0,\ldots,d-1$;  (2) for each $i$ there is some $j\neq i$ such that if the $j$-th row has nonzero entries at $(j,j_1)$ and $(j,j_2)$ then $i-i_1 = j-j_1$, $i-i_2 = j- j_2$.  The reason for imposing the two conditions will be made clear later on.

We find $D_{\alpha}, D_{\beta}\in \hat{A}[\ell]$ along with
nontrivial $\lambda, \mu \in \End (\hat{A}[\ell])$ such that $\lambda (D_{\beta}) = D_{\alpha}$, and $\mu (D_{\beta})= 0$ on $\hat{A}$ and
$\hat{e}(D_{\alpha}, D_{\beta}) \neq 1$.

Moreover $\lambda$ and $\mu$ can be specified as a linear sum over $\varphi_i$.  So,
$\lambda = a_0+\sum_i a_i \varphi_i$ and $\mu = b_0+\sum_i b_i \varphi_i$ with $a_i, b_i \in \F_{\ell}$.

Using the descent basis and the maps $\rho$ defined by the basis and
the matrices $M_i$'s,
we can construct the two points together with $\lambda$ and $\mu$ from points on $A[\ell]$.  One way to do this is as follows.

Without loss of generality assume $\mu_{\ell}\subset K$.
Find $a,b\in A(K)[\ell]$ such that $e_{\ell} (a,b)\neq 1$.
Choose random $x_i,y_i\in\F_{\ell}$ and let $D_{\beta}\in\hat{A}[\ell]$ such that $D_{\beta}$ corresponds to $V=(x_i a+ y_i b)_{i=0}^{d-1}\in A[\ell]^d$.

Let $v_1 = (x_i)_{i=0}^{d-1}$ and $v_2 = (y_i)_{i=0}^{d-1}$.

Let $z_i \in \F_{\ell}$ for $i=0,\ldots,N_1$.  Then
$\sum_i z_i \varphi_i (D_{\beta}) =0$ if and only if $\sum_i z_i M_i V =0$ if
$\sum_i z_i M_i v_1 = 0 \mod \ell$ and $\sum_i z_i M_i v_2 = 0 \mod \ell$.

This leads to $2d$ linear equations over $\F_{\ell}$ in $d^{O(1)}$ variables.  We expect there to be many solutions.
Choose one such solution and set $\mu = \sum_i z_i \varphi_i$.  Then
$\mu (D_{\beta}) = 0$.

Choose random $c_i\in\F_{\ell}$ such that let $\sum_i c_i M_i (V) = (x'_i a + y'_i b)_{i=0}^{d-1}$ then
\[\prod_i e_{\ell}( (x'_i a + y'_i b ), (x_i a + y_i b ) ) \neq 1.\]

Set $\lambda = \sum_i c_i \varphi_i$.

Set $D_{\alpha} = \lambda (D_{\beta})$.  Then $D_{\alpha}$ corresponds to $(D'_i)\in A[\ell]^d$ where
$D'_i = x'_i a + y'_i b$, and
\[\hat{e} (D_{\alpha}, D_{\beta}) = \prod_i e_{\ell}( (x'_i a + y'_i b ), (x_i a + y_i b ) ) \neq 1.\]

Let $G_1$ and $G_2$ be respectively the cyclic groups generated by $D_{\alpha}$ and $D_{\beta}$.

Let $\Lambda$ be the $\F_{\ell}$ associative non-commutative algebra generated by a set $\Sigma$ of $N_1$ variables $z_1,\ldots,z_{N_1}$.

Let $\phi$ be the morphism of algebra from $\Lambda$ to $\End(\hat{A}[\ell])$ such that $\phi(z_i)=\varphi_i$ for all $i$.  Then $\phi$ defines an action of $\Lambda$ on $\hat{A}[\ell]$.

Let $f_{\lambda} = a_0+\sum_i a_i z_i$ and $f_{\mu} = b_0+\sum_i b_i z_i$, so that $\phi (f_{\lambda}) = \lambda$ and $\phi (f_{\mu}) = \mu$.

For $n\in\Z_{\ge 0}$, let $\Lambda_n$ denote the submodule of $\Lambda$ spanned by monomials over $\Sigma$ of degree no greater than $n$.

Set a bound $N=O(d)$ and let $S=\{f_{\lambda}\}\cup \{ wf_{\mu}: w$ is a monomial over $\Sigma$ of degree less than $N\}$.

Let $U$ be the submodule of $\Lambda$ spanned by $S$.
Let $G_3 = U_1/U$ with $1+U$ as the generator.

For $z\in\F_{\ell}$, $z+U\in G_3$ is encoded by a sparse random representative  $\gamma\in z+U\subset U_1\subset\Lambda_N$.
More precisely,  we randomly select $t=d^{O(1)}$ elements $w_i\in S$ and random $a_i\in\F_{\ell}$, then compute $\gamma=z+\sum_i a_i w_i$ as an element in $\Lambda_N$.
Then $\gamma\in \Lambda_N$ is an encoding of $z+U\in G_3$.

The trilinear map $G_1\times G_2 \times G_3 \to \mu_{\ell}$ sends
$(xD_{\alpha},yD_{\beta},z+U)$ to $\zeta^{xyz}$ where $\zeta=\hat{e}(D_{\alpha},D_{\beta})$.
Suppose $z+U$ is represented by $\gamma\in z+U$.  Then
\[ \hat{e}(xD_{\alpha},\phi(\gamma)(yD_{\beta}))=\hat{e} (xD_{\alpha},zyD_{\beta})=\zeta^{xyz}.\]

The sparsity constraint is to make sure that the map $\gamma$ can be efficiently executed,so that the trilinear map can be efficiently computed.

\subsection{Specifying the trilinear map}
Fix a public basis $\theta_1,\ldots,\theta_d$ of $K/k$, a private basis $u_1,\ldots,u_d$ of $K/k$, and another private basis $u'_1,\ldots,u'_d$ of $K/k$.
The private basis $u_1,\ldots,u_d$ and the associated descent table (which is the multiplication table for the basis) as well as the conversion table $(c_{ij})$, so that $u_i = \sum_{j=1}^d c_{ij} \theta_j$, with $c_{ij}\in k$ for $1\le i,j \le d$, are all hidden.  The second private basis and the associated descent table and conversion table are hidden likewise.
We keep $A$ secret as well.

Let $\delta$ denote the basic descent map $\bar{k}^d \to \bar{k}$ with respect to $u_1,\ldots,u_d$, and $\rho$ the bijective linear map $\bar{k}^d \to \bar{k}^d$ determined by $\delta$.
So
$\delta(x) = \sum_{i=1}^d x_i u_i$ and
$\rho (x) = (\delta^{\sigma^i} (x) )_{i=0}^{d-1}$ for $x=(x_i)_{i=1}^d\in \bar{k}^d$.

Let $\delta'$ denote the basic descent map $\bar{k}^d \to \bar{k}$ with respect to $u'_1,\ldots,u'_d$, and $\rho'$ the bijective linear map $\bar{k}^d \to \bar{k}^d$ determined by $\delta'$.

Let $\hat{A}$ denote the descent of $A$ with respect to the basis $u_1,\ldots,u_d$.

Let $\hat{A}'$ denote the descent of $A$ with respect to the basis $u'_1,\ldots,u'_d$.

We publish the following
\begin{enumerate}
\item  $D'_{\alpha}$ and $D_{\beta}$ where $D'_{\alpha}$ is the image of $D_{\alpha}$ under the natural isomorphism
between $\hat{A}$ and $\hat{A}'$ determined by $\rho'^{-1}\rho$,
\item the program for computing the descent $\hat{m}$ of the addition $m$ on $\hat{A}$,
the program for computing the descent $\hat{m}'$ of the addition $m$ on $\hat{A}'$,
\item  the programs for computing $\varphi_i$, $i=1,\ldots,N_1$,
\item  $f_{\lambda}$ and $f_{\mu}$,
\item the program for computing $\hat{e}:\hat{A}'[\ell]\times \hat{A}[\ell]\to\mu_{\ell}$, including additional programs for efficient computation of $\hat{e}$.
\end{enumerate}

The points $D'_{\alpha}$ and $D_{\beta}$, maps $\varphi_i$, $i=1,\ldots,N_1$, and additional programs mentioned above are defined over $K$.  The polynomials that we use to specify these programs have coefficients in $K$ written out in the public basis $\theta_1,\ldots,\theta_d$.

In specifying programs for the descent addition morphism $\hat{m}$, $\hat{m}'$ and $\varphi_i$'s, we want to make sure that the descent basis remain secret.  We will discuss how this can be done for maps on descent varieties in general in the next section.

The encoding of $x\in\F_{\ell}$ and $y\in\F_{\ell}$ by $x D'_{\alpha}$ (resp. $y D_{\beta}$) requires $O(\log \ell)$ applications of $\hat{m}'$ (resp. $\hat{m}$).

The encoding of $z\in\F_{\ell}$ by a $t=d^{O(1)}$ sparse element $\gamma\in U_1\subset \Lambda_N$ is of length $d^{O(1)}$. the computation of $\phi(\gamma) (y D_{\beta})$ requires $d^{O(1)}$ applications of $\varphi_i$'s and $\hat{m}$.

The cyclic groups $G_i$, $i=1,2,3$, are DDH-groups constructed with the two secret descent bases as trapdoor. The reason for using a different descent basis in constructing $G_1$ is so that the efficient pairing $G_1 \times G_2 \to \mu_{\ell}$ cannot be used to define a self pairing on $G_1$ or $G_2$.   We note that if the two secret descent bases were identical then the published pairing $\hat{e}$ together with some $\varphi_i$ can be used to induce self pairing on $G_1$.  Namely if $\hat{e} (D_{\alpha}, \varphi_i (D_{\alpha}))\neq 1$, then we have an efficiently computable pairing $G_1\times G_1 \to \mu_{\ell}$, hence $G_1$ would not satisfy DDH assumption.  Similar observation applies to $G_2$.  As for $G_3$, neither the pairing $\hat{e}$ nor the trilinear map naturally induce a self pairing on the group.

\subsection{The discrete logarithm problem on $G_3$}
One way to think about the discrete logarithm problem on $G_3$, in the setting described above, is that it is a discrete logarithm problem with a trapdoor and efficient public zero (identity) testing.  The trapdoor is the secret descent basis, and public zero testing can be efficiently done through the specified trilinear map.

First let us consider the generic discrete logarithm problem, where the specification of $G_1$, $G_2$ and the program for computing the trilinear map is removed.
The $\F_{\ell}$-dimension of $U$ is exponential in $d$ given that the cardinality of $S$ is exponential in $d$.
A linear attack would naturally require generating exponentially many random elements of $U$ in order to construct a basis for $U$ and solve the discrete logarithm problem on $U_1/U$ by reduction to linear algebra in $\Lambda$.

We note that even checking whether an element of $U$ is in $U$, that is to say zero testing on elements of $G_3$, already seems difficult.

Next consider $U$ as constructed before, except that $f_{\lambda}$ and $f_{\mu}$ are more generally random low degree elements in $\Lambda$. Let $Mat_d (\F_{\ell})$ denote the algebra of $d$ by $d$ matrices over $\F_{\ell}$, and
consider the morphism $\psi: \Lambda \to Mat_d (\F_{\ell})$ determined by $\psi ( z_i) = M_i$ for $i=1,\ldots,N_1$.

The map $\psi$ can be regarded as a trapdoor for solving the discrete logarithm problem on $G_3$. If we know $\psi$, then $M_i = \psi (z_i)$ can be determined, and the discrete logarithm problem on $U_1/U$ is reduced to linear algebra in $Mat_d (\F_{\ell})$, an $\F_{\ell}$ vector space of dimension $d^{O(1)}$.

Therefore, if we consider $f_{\lambda}$ and $f_{\mu}$ as the public key, and $\psi$ as the secret key, and consider the encoding of $z\in\F_{\ell}$ is by a $t=d^{O(1)}$ sparse element $\gamma\in U_1\subset \Lambda_N$, then we have a public key encryption scheme.
Decoding is easy if we have the secret key $\psi$, otherwise we are faced with the generic discrete logarithm problem on $G_3 = U_1/U$.

However we do not have efficient zero testing on $G_3$ yet.

In the trilinear map setting involving the descent abelian variety $\hat{A}$, additional information is provided, such as programs specified for computing $\varphi_i$'s.   If the descent basis is known then $\rho$ is known, and $M_i$ can be determined(see the diagram below)
\[
\begin{array}{llll}
\hat{A}[\ell] & \stackrel{\rho}{\to} & \prod_{\sigma} A^{\sigma} [\ell]\cong  & A[\ell]^d\\
\downarrow \varphi_{i} & &   & \downarrow M_i\\
\hat{A}[\ell] & \stackrel{\rho}{\to} & \prod_{\sigma} A^{\sigma} [\ell]\cong  & A[\ell]^d

\end{array}
\]
as we consider the action of $\varphi_i$ on $D_{\alpha}$ or $D_{\beta}$.
Consequently the map $\psi$ is determined.  Moreover with the efficient program for the trilinear map, efficient public zero testing in $G_3$ can be done.

We also note that the discrete logarithm problem on $G_i$ for $i=1,2,3$ is reduced to the discrete logarithm problem on $\mu_{\ell}\subset k(\mu_{\ell})$.

We can modify $U$ and $U_1$ to make the discrete logarithm problem on $U_1/U$ potentially more difficult.  For example, we determine for each $i,j$ pair with $i\neq j$ a relation $M_i M_j = r_{ij} M_j M_i + \sum_k a_{ijk}M_k$,  with $r_{ij}, a_{ijk}\in\F_{\ell}$. Then publish the equality $\varphi_i \varphi_j = r_{ij} \varphi_j \varphi_i + L_{ij}$, where $L_{ij}=\sum_k a_{ijk}\varphi_k$.
The element
 $R_{ij}=\varphi_i \varphi_j - r_{ij} \varphi_j \varphi_i - L_{ij}$ will be included in $U$, as well as
 $w_1 R_{ij} w_2$ for any monomials $w_1,w_2$ over $\Sigma$ such that $w_1w_2$ is of degree $O(d)$.

Let $w$ be a monomial of degree $O(d)$ over $\Sigma$ containing $\varphi_i \varphi_j$ so that $w=w_1 \varphi_i \varphi_j w_2$ where $w_1$ and $w_2$ are monomials over $\Sigma$.  Then $w\equiv w_1 (r_{ij} \varphi_j \varphi_i + L_{ij}) w_2\mod U$.
Suppose $f=aw+f_2\in U_1$ with $a\in\F_{\ell}$.  Then modifying $f$ to $f'= aw_1 (r_{ij} \varphi_j \varphi_i + L_{ij}) w_2 +f_2$ is called a {\em switch}.

Now to encode $z$ by a random element $\gamma\in z+U$, first choose as before a random sparse linear expression $\gamma'$ over $S$, where the number of nonzero terms is $d^{O(1)}$, then randomly perform $d^{O(1)}$ switches on $\gamma'$ to obtain $\gamma$.

Below we discuss how $\varphi_i$ can be specified algebraically and why the following two conditions are imposed in choosing $M_i$'s: (1) there are exactly two nonzero entries $(i,i_1)$ and $(i,i_2)$ for row $i$, for $i=0,\ldots,d-1$;  (2) for each $i$ there is some $j\neq i$ such that if the $j$-th row has nonzero entries at $(j,j_1)$ and $(j,j_2)$ then $i-i_1 = j-j_1$, $i-i_2 = j- j_2$..

Recall that through the isomorphism $A[\ell]\to A_i [\ell]: \alpha\to \sigma^i (\alpha)$, we have an isomorphism $\Delta$ between $\hat{A}[\ell]$ and $d$-fold product of $A[\ell]$ identifying $D\in\hat{A}[\ell]$ with $(D_i)_{i=0}^{d-1}$ where $D_i=\sigma^{-i}(\delta^{\sigma^i} D)=\delta \sigma^{-i} D$. With the identification
a $d\times d$ matrix $M=(a_{ij})$ with $a_{ij}\in\F_{\ell}$ define an element $\phi_M \in \End (\hat{A}[\ell])$, so that
if $D\in \hat{A}[\ell]$ is identified with $\alpha=(D_i)_{i=0}^{d-1}$ through $\Delta$ then
$\phi_M (D)$ is identified with $M \alpha$.

Consider a matrix $M$ in the set.  We have $\varphi_M (D) = \rho^{-1}v$ where $v= (v_i)_{i=0}^{d-1}$ with $v_i=\sigma^i m(\delta \sigma^{-i_1} D, \delta \sigma^{-i_2} D)$.  If $D\in\bar{k}^n$ then $v$ can be regarded as a $d$ by $n$ matrix where the $i$-th row is $v_i$.

More precisely, let $F(X,Y)$ be the $n$-vector of polynomials such that $F(D_1,D_2)=m (\delta D_1, \delta D_2)$ on an affine piece of $\hat{A}$ contained in $\bar{k}^n$.  Note that $F(D_1,D_2) = \tilde{m} (D_1,D_2)$.
Let $F'_i (X, Y) = F^{\sigma^i} (X^{q^{d-i_1 +i}}, Y^{q^{d-i_2 +i}} )$, which is obtained by substituting in $F^{\sigma^i} (X,Y)$ each variable $x$ in the $X$ part by $x^{q^{d-i_1 +i}}$ and each variable $y$ in the $Y$ part by $y^{q^{d-i_2 +i}}$.  Then $v_i=F'_i (D,D) = \sigma^i m(\delta \sigma^{-i_1} D, \delta \sigma^{-i_2} D)$  for $D\in\hat{A}(K)$.

We see that for $D\in\hat{A}(K)$, $\varphi_M (D) = (g_j (D))_{j=0}^{d-1}$ where $g_j (D) = \sum_i \alpha_{ji} F'_i (D,D)$ and
$\rho^{-1} = (\alpha_{ji})$ with $0\le j,i \le d-1$.
So the map $\varphi_M$ can be specified by polynomials in $q^j$-th powers of variables for various $j$.

Below we argue heuristically that from the specification of $\varphi_M$, it is hard to extract information on either $\rho$ or $M$.

The value of $i_1$ and $i_2$ for $i$ can be read off from the $i$-th entry of $v$.  However $v$ is not revealed, but $u$ is where $u=\rho^{-1} v$.
Consider the $j$-th entry of $u$ for example, which is $g_j (D) = \sum_i \alpha_{ji} F'_i (D,D)$.  The terms in the specified polynomial $g_j$
appear according to a monomial ordering.  The set of $(a,b)$, where $a=d-i_1+i$ and $b=d-i_2+i$ for some row $i$, may still be recognized from the exponents of variables in $g_j$.
However from each $(a,b)$ one cannot determine $i$, $i_1$ and $i_2$ such that $a=d-i_1+i$ and $b=d-i_2+i$.

We see that, at least heuristically, it is difficult to determine the matrix $M$ from the exponents of the polynomials in the product $u=\rho^{-1}v$ that gives the specification of $\varphi_M$.

Write $\varphi_M (D) =\rho^{-1} v=(\rho^{-1} v^{(j)})_{j=1}^n$ where $v^{(j)}$ is the $j$-th column of $v$.
Let $u_j=\rho^{-1} v^{(j)}$.
We can write $u_j = \sum_{\alpha} \nu_{\alpha} t_{\alpha}$ where $t_{\alpha}$ is a monomial and $\nu_{\alpha}\in\bar{k}^d$ is its coefficient vector.
Suppose that a monomial $t_{\alpha}$ appears in the $i$-th entry of $v^{(j)}$ with coefficient $a_i$ for $i=0,\ldots,d-1$.  Then the coefficient vector of $t_{\alpha}$ in $u_j$ is $\sum_i a_i C_i$ where $C_i$ is the $i$-th column of $\rho^{-1}$.
In particular if
there is a monomial $t_{\alpha}$ that appears only in the $i$-th entry of $v^{(j)}$, then the coefficient vector of $t_{\alpha}$ in $u_j$ is $C_i$, the $i$-th column of $\rho^{-1}$, up to a scalar multiple.    However this situation does not arise since we require that there is some other row $i'$ of the matrix $M$ such that if the $i'$-th row has nonzero entries at $(i',i'_1)$ and $(i',i'_2)$ then $i-i_1 = i'-i'_1$, $i-i_2 = i'- i'_2$.  This means that the same monomial appears in the $i'$-th entry of $v^{(j)}$ as well.

We see that, at least heuristically, it is difficult to extract information on $\rho^{-1}$ from the product $u=\rho^{-1}v$ that gives the specification of $\varphi_M$.

In \S~\ref{jac} we present a concrete construction of trilinear maps involving the jacobians of hyperelliptic curves.

\section{Weil descent for secrecy}
\label{secrecy}
In specifying programs for the descent addition morphism $\hat{m}$, we want to make sure that the descent basis remain secret.  In this section we discuss how this can be done for descent maps on descent varieties in general.

When we refer to Weil descent we mean descent with respect to the private basis $u_1,\ldots,u_d$.

\subsection{Global descent}

A {\em term} $T$ with coefficient $a$ is of the form $am$ where $a$ is a constant and $m$ is a monomial. Call a term $T$ {\em vital} if it is of degree greater than 1 or of the form $ax_i$ where $K=k(a)$.

The {\em support} of a polynomial is the set of monomials that appear in the polynomial with nonzero coefficient.

For $a\in K$, let $\Gamma_a = (\gamma_{ij})$ be the $d$ by $d$ matrix in $Gl_d (k)$ such that
$au_i = \sum_{j=1}^d \gamma_{ij} u_j$.  Note that the fraction of $\Gamma\in Gl_d (k)$ such that $\Gamma=\Gamma_a^t$ for some $a\in K$ is in roughly
$\frac{|k|^d}{|k|^{d^2}}$, which is negligible.

\begin{theorem}
\label{gld}
\begin{enumerate}
\item  Suppose $F\in R$ contains a vital term.  Then given $\hat{F}$ one can efficiently uncover the descent basis.
\item Given $(f_i)_{i=1}^d$ with $f_i\in\hat{R}$ and the descent basis, one can efficiently check if there is some $F\in R$ such that $\hat{F}=(f_i)_{i=1}^d$.
\item  Suppose $F\in R$ and $\Gamma\in Gl_d (k)$.
If $\Gamma\hat{F}$ contains the global descent of a nonconstant term, then $\Gamma = \Gamma_a$ for some $a\in K$.
If $\Gamma\hat{F}=\hat{G}$ for some $G\in R$.  Then $G=aF$ for some $a\in K$ and $\Gamma = \Gamma^t_a$.  Consequently, the fraction of $\Gamma\in Gl_d(k)$ such that $\Gamma \hat{F}$ contains a global descent of a nonconstant term is negligible.
\end{enumerate}
\end{theorem}

\ \\{\bf Proof}
(1)
Write $F$ as the sum of terms $F=\sum T_i$.  Then $\hat{F}=\sum_i \hat{T_i}$.   From $\hat{F}$ we can read off $\hat{T_i}$ easily since $\hat{T_i}$ have disjoint supports, each determined completely by the corresponding monomial in $T_i$.  So it is enough to consider the case where $F$ is a vital term $T$.

Suppose $F=T$ is a vital term and for simplicity suppose  $T=a x_1\ldots x_r$ for some $r\ge 1$, where either $r > 1$ or $r=1$ and $K = k(a)$.  Below we discuss how $u_1,\ldots,u_d$ can be uncovered from $\hat{T}$.

Suppose $\tilde{T}= \sum_{i=1}^d h_i u_i$.
Set $b=au_2\ldots u_r$ if $r > 1$.  Then
\[ \widetilde{bx_1} = \sum_{i=1}^d h_i (\hat{x}_1, \hat{u}_2,\ldots,\hat{u}_r) u_i .\]
So $\widehat{bx_1}$ can be obtained from $\hat{T}$.
It is likely that $b$ generates $K$ over $k$, in which case from $\widehat{b u_j}$, $j=1,\ldots,d$, we compute the irreducible polynomial for $b$, and determine $b$ up to Galois conjugates.

 Evaluating $\widehat{bx_1}$ at $\hat{x}_1 =\widehat{bu_j}$ we obtain $\widehat{b^2 u_j}$.  Iterating we obtain $\widehat{b^i u_j}$ for $i=1,\ldots,d-1$.  From these and the irreducible polynomial of $b$ we can determine $u_j$ as a polynomial expression in $b$.  In this fashion the basis $u_1,\ldots,u_d$ can be uncovered.

\ \\(2) Given a $d$-tuple of polynomials $(f_i)_{i=1}^d$ with $f_i \in \hat{R}$, we can verify whether the tuple contains the descent of some polynomial with the help of descent tables.   From the terms of the $d$ polynomials we can determine a set of monomials $m_1, \ldots, m_t$ in $R$ so that the support of each $f_i$ is contained in the union of the supports of $\hat{m}_1,\ldots,\hat{m}_t$.  Write $f_i = \sum_{j=1}^t f_i^{(j)}$ where the support of $f_i^{(j)}$ is contained in the support of $\hat{m}_j$, so that $(f_i)_{i=1}^d = \sum_{j=1}^t (f_i^{(j)})_{i=1}^d$.
If $(f_i)_{i=1}^d$ contains the descent of some nonconstant term, then
\[ (f_i^{(j)})_{i=1}^d = \widehat{a_j m_j}\] for some $a_j\in K$.

We are reduced to checking if a $d$-tuple is $\widehat{am}$ for some $a\in K$, given the tuple and monomial $m$.

Suppose $m=x_1^{e_1}\ldots x_n^{e_n}$ and $a=\sum_{i=1}^d a_i u_i$.  Suppose $m$ is of degree $d_m$.  Then

\begin{eqnarray*}
\widetilde{am} & = & (\sum_{i=1}^d a_i u_i) (\sum_{i=1}^d x_{1i} u_i)^{e_1}\ldots(\sum_{i=1}^d x_{ni} u_i)^{e_n}\\
 & = & \sum_{1\le i_0,i_1,\ldots,i_n\le d} a_{i_0} x_{1i_1}\ldots x_{ni_{d_m}} u_{i_0}u_{i_1}\ldots u_{i_{d_m}} \\
 & = & \sum_j \sum_{i_1,\ldots,i_{d_m}} \sum_{i_0} a_{i_0} \delta_{i_0,i_1,\ldots,i_{d_m}, j} x_{1i_1}\ldots x_{ni_{d_m}} u_j
\end{eqnarray*}

By comparing coefficients with the $d$ polynomials we get a system of $d$ linear equations in the unknown $a_i$, $i=1,\ldots,d$.  The $d$ polynomials form a descent if and only if the system has a solution.

\ \\(3)
Consider a non-constant term $T \in R$.  Suppose $\tilde{T}=\sum_{i=1}^d f_i u_i$.  Then for $a\in K$,
\[ \widetilde{aT} = \sum_{i=1}^d f_i au_i=\sum_j \sum_i f_i \gamma_{ij} u_j.\]
Hence
\[ \widehat{aT} = \Gamma_a^t \hat{T}.\]

It is easy to see that $\{ \hat{T} (\hat{\alpha}) : \alpha\in K^n\}$ contains $d$ linearly independent vectors since $\widehat{T(\alpha)} = \hat{T}(\hat{\alpha})$.  Hence for $\Gamma \in Gl_d (k)$, $\Gamma \hat{T} = \hat{T}$ if and only if $\Gamma$ is the identity matrix.  It follows that $\Gamma \hat{T} = \widehat{aT}$ if and only if $\Gamma=\Gamma^t_a$.

Now let $F=\sum_i T_i$ where $T_i$ is a term.  Let $\Gamma \in Gl_d (k)$.  Then $\hat{F}=\sum_i \hat{T}_i$ and
$\Gamma\hat{F} = \sum_i \Gamma \hat{T}_i$.  If $\Gamma \hat{F}$ contains a nontrivial global descent, then
$\Gamma\hat{T}_i$ is a global descent for some $i$ where $T_i$ is a non-constant term.  This implies $\Gamma\hat{T}_i = \widehat{aT}$ for some $a\in K$.  It follows that $\Gamma =\Gamma^t_a$.

In particular if $\Gamma \hat{F}=\hat{G}$ then $G=aF$ for some $a\in K$ and $\Gamma = \Gamma_a$.

\subsection{Specifying maps on descent varieties}
Suppose $V=Z(F_1,\ldots,F_m)$, the algebraic set defined by the zeroes of $F_1,\ldots,F_m \in R$.
We consider the problem of specifying maps on $\hat{V}$ in such a way that the descent basis remains secret.

Suppose a map $\varphi: V(\bar{k}) \to \bar{k}$ can be defined by the restriction of a polynomial $H\in R$ to $V$.
Then $\hat{\varphi}$ can be defined by the restriction of $\hat{H} = (h_i)_{i=1}^d$ to $\hat{V}$, with $h_i \in \hat{R}$ with coefficients in $k$.  However as discussed above the global descent $(h_i)_{i=1}^d$ can likely be used to uncover the descent basis. Therefore we cannot specify $\hat{\varphi}$ by $(H_i)_{i=1}^d$.  Instead we will specify $\hat{\varphi}$ by some $(h'_i)_{i=1}^d$ where $h'_i = h_i +g_i$ with $g_i\in\hat{R}$ and $g_i$ vanishes on $\hat{V}$, so that $(h'_i )_{i=1}^d$ contains no global descent.  Simply put we want $h'_i = h_i \mod I(\hat{V})$ such that $(h'_i)_{i=1}^d$ contains  no global descent.

Similarly if $\tilde{\varphi}$ can be defined by the restriction of $\sum_{i=1}^d f_i \theta_i$ with $f_i\in\hat{R}$, we would like to make sure that $(f_i)_{i=1}^d$ does not contain any global descent.

More generally we consider the following problem.  Given $(f_i)_{i=1}^d$ with $f_i \in\hat{R}$, we want to construct
$f'_i$, $i=1,\ldots,d$, such that $f'_i = f_i \mod I(\hat{V})$ and $(f'_i)_{i=1}^d$ contains  no global descent.

 From the terms of the $d$ polynomials we can determine a set of monomials $m_1, \ldots, m_t$ in $R$ so that the support of each $f_i$ is contained in the union of the supports of $\hat{m}_1,\ldots,\hat{m}_t$.
 We can rewrite $(f_i)_{i=1}^d$ as $\sum_{i=1}^t H_i$ with $H_i \in\hat{R}^d$ and the support of $H_i\in \hat{R}^d$ is a subset of the support of  $\hat{m}_i$.  For each $i$ if $H_i$ is not a global descent, put $H'_i = H_i$.  If $H_i$ is a global descent, then $H_i =\hat{T}$ where $T=am_i$ for some $a\in K$.
We
choose a polynomial $F$ that vanishes on $V$ such that $m_i$ appears in $F$ with nonzero constant.  (For example if $F_1$ has a nonzero constant term, then we can take $F=bm_i F_1$ with random nonzero $b\in K$.)

Let $\Gamma$ be randomly chosen from $Gl_d (k)$.  Then by Theorem~\ref{gld}, $\Gamma \hat{F}$ is most likely not a global descent.   In that case  $\hat{T}+\Gamma\hat{F}$ does not contain a global descent either.  Otherwise $\hat{T} + \Gamma\widehat{bm_i} = \widehat{cm_i}$ for some $c\in K$, but then $\Gamma\widehat{bm}=\widehat{cm_i}-\widehat{am_i} = \widehat{(c-a)m_i}$, and we have a contradiction.

Put $H'_i = H_i +\Gamma\hat{F}$. If we write $\sum_i H'_i = (g_i)_{i=1}^d$ with $g_i \in\hat{R}$, then $g_i = f_i \mod I(\hat{V})$.

Suppose a map $\varphi: V(\bar{k}) \to \bar{k}$ can be defined by the restriction of a polynomial $H\in R$ to $V$.
Then $\hat{\varphi}$ can be defined by the restriction of $\hat{H}\in\hat{R}^d$ to $\hat{V}$.  Let $H=\sum_i a_i m_i$ where $a_i\in K$ and $m_i$ is a monomial.   Then
$\hat{H} = \sum_i H_i$ where $H_i = \widehat{a_i m_i}$.
Let $\hat{H}=(h_i)_{i=1}^d$ with $h_i \in \hat{R}$.
After the above procedure is applied to all $H_i$, we obtain some $(h'_i)_{i=1}^d$ with $h'_i\in\hat{R}$ and
$h'_i = h_i \mod I(\hat{V})$.
Moreover if we write $(h'_i)_{i=1}^d = \sum_i H'_i$ where $H'_i \in \hat{R}^d$.  Then each $H'_i$ is of the form
$H'_i=\sum_j \Gamma_{ij}) \hat{m}_i$ where $m_i$ is a monomial and $\Gamma_{ij}\in Gl_d (k)$ and all but at most one $\Gamma_{ij}$ are random elements in $Gl_d (k)$.
It is unlikely that $\sum_j \Gamma_{ij} = \Gamma^t_a$ for some $a\in K$.  Consequently it is unlikely that $H'_i = \widehat{am_i}$ for some $a\in K$. That is, $(h'_i)_{i=1}^d$ is unlikely to contain a global descent.

Now consider the map $\tilde{\varphi}$.  Since
\[ \tilde{\varphi} (\hat{\alpha}) = \sum_{i=1}^d \hat{\varphi}_i (\hat{\alpha}) u_i
=\sum_{i=1}^d \psi_i (\hat{\alpha}) \theta_i\]
where $\psi_i (\hat{\alpha}) = \sum_{j=1}^d \hat{\varphi}_j (\hat{\alpha})c_{ji}$,
$u_i = \sum_{j=1}^d c_{ij} \theta_j$, with $c_{ij}\in k$ for $1\le i,j \le d$.
Let $\Gamma=(c_{ij})$.  Then $\Gamma\in Gl_d(k)$.
If $\varphi$ can be defined as the restriction of $H\in R$ to $V$, then $\tilde{\varphi}$ can be defined as the restriction of $\Gamma\hat{H}$ to $\hat{V}$.
We can make sure that there is no $a\in K$ such that $a (u_i)_{i=1}^d = (\theta_i)_{i=1}^d$, hence
$\Gamma\neq \Gamma_a^t$ for any $a\in K$, so by Theorem~\ref{gld}, $\Gamma\hat{H}$ contains no global descent.

In the same way we can specify the descent of a map $V(\bar{k}) \to V(\bar{k})$ to $\hat{V}(\bar{k}) \to \hat{V} (\bar{k})$ such that the specification does not contain the global descent of any nontrivial term.

\subsection{Linear-term attack}
\label{analysis}

Suppose the descent $\hat{\varphi}$ of a map $\varphi: V\to V$ defined over $K$ is specified properly so that the description of $\hat{\varphi}$ contains  no global descent. Suppose one point on $\hat{V}$ is given. Then  starting with the given point, one can repeatedly apply the descent map $\hat{\varphi}$ to obtain more points on $\hat{V}$.   Heuristically speaking we may consider these points as random sampling of $\hat{V} (\bar{k})$.
An interesting question from the attacker's perspective is:  can $V$ be efficiently uncovered after sampling polynomially many points of $\hat{V}$?

Most points on $\hat{V}$ are not descent points.   If a descent point $\hat{\alpha}$ of some $\alpha \in V(\bar{k})$ is known and $\alpha$ is not $K$-rational, then a lot of information can be revealed about the descent basis from $\hat{\alpha}$.  To see this let $\hat{\alpha}=(\beta_i)_{i=1}^d$.  Then $\alpha^{\sigma^i}=\sum_{j=1}^d \beta_j u_j^{\sigma^i}$, for $i=0,\ldots,d-1$.
So $\alpha= \sum_{j=1}^d \beta_j^{\sigma^{-i}} u_j$.
Since $\alpha = \sum_{j=1}^d \beta_j u_j$, we have  $\sum_{j=1}^d (\beta_j^{\sigma^{-i}} - \beta_j) u_j =0$.
When $\alpha$ is not $K$-rational, $\beta_j$ may not be fixed by $\sigma$, and we have a non-trivial linear condition on
$u_1,\ldots,u_d$.

In our situation we can assume that only descent points of $K$-rational points are revealed through computation.

Suppose neither global descents nor descent points (embedded image of points on $V$) is revealed.
It is still an interesting question to come up with a strategy to uncover the descent basis from the sampled points on $\hat{V}$.
Once the basis is uncovered, we can map the sampled points back to obtain points on $V$.  Thus we can likely recover $V$.

To uncover the descent basis, one strategy is to form a linear space of polynomials with bounded support that vanish at all the sampled points, and try to find from the linear space a global descent. As discussed before, once we have a global descent we are likely to uncover the descent basis.

In general suppose $S$ is a finite set of monomials.
Let $L_S$ be the linear space of polynomials in the ideal of $V$ with support bounded by $S$.
Let $L_{\hat{S}}$ be the linear space of polynomials in the ideal of $\hat{V}$ with support bounded by $\hat{S}$.     If $F\in L_S$, then $L_{\hat{S}}$ contains all $d$ polynomials in $\hat{F}$.  In addition for every $\Gamma\in Gl_d (k)$, the $d$-tuple of polynomials in $\Gamma \hat{F}$ are all in $L_{\hat{S}}$ as well.
By Lemma~\ref{gld} we know that the fraction of $\Gamma\in Gl_d (k)$ such that $\Gamma=\Gamma_a^t$ for some $a\in K$ is in roughly
$\frac{|k|^d}{|k|^{d^2}}$, which is negligible.
Therefore, to
dig out a $d$-tuple of polynomials that form a global descent a very targeted search is required.
We assume heuristically that after sufficiently many points are sampled  $L_{\hat{S}}$ is the linear space of polynomials in $\hat{R}$ with support bounded by $\hat{S}$ that vanishes at all the sample points.

One special case where this is possible is when there is some linear $F$ that vanishes on all $\alpha$ such that $\hat{\alpha}$ is a sampled point.  In this case we may as well consider the minimal linear variety that contains all such $\alpha$ and its descent.  For simplicity assume the minimal linear variety is defined by one linear polynomial $F$. Then $V=Z(F)$ is of dimension $m-1$ and $\hat{V}$ is a linear variety of dimension $(m-1)d$ defined by the $d$ linear polynomials in $\hat{F}$.  Assume without loss of generality that the coefficient of $x_1$ in $F$ is 1.  Let $\hat{F}=(f_i)_{i=1}^d$.  Then the coefficient of $y_{1j}$ in $f_i$ is all 0 except for $j=i$.  Hence a targeted search for $f_i$ is possible.  More exactly
we set $S=\{ x_1,\ldots, x_m\}$ and correspondingly $\hat{S}=\{ y_{ij}: 1\le i\le m, 1\le j \le d\}$.  We see that   $L_{\hat{S}}$ is of dimension $d$ with $\hat{F}$ as a special basis that is easy to identify: $f_i$ can be obtained by further restrictions that the coefficients for $y_{1j}$ is 0 for $j\neq i$.  These $d-1$ additional linear conditions likely allows us to extract $f_i$.

The linear case is special in that the conditions for the desired descent can  be described without reference to the descent table.  The above attack can extend to the case when $V$ is defined by a polynomial $F$ that contains a linear term, if $L_{\hat{S}}$ is of dimension $d$ where $S$ is the support of $F$ (in general $\dim L_{\hat{S}}\ge d$).  We may again assume without loss of generality that the coefficient of $x_1$ in $F$ is 1, then $L_{\hat{S}}$ has $\hat{F}$ as a special basis that is easy to identify.  We call this the {\em linear term} attack.  This analysis suggests that the case where $V$ is a hypersurface is a relatively weak case.

 Suppose $V\subset \bar{k}^n$ is a variety of dimension $n-g$ defined by $g$ polynomials in $R$. Let $S$ be the support of the defining set of polynomials. As before assume that random sampling of points on $\hat{V}$ is available,  then the linear term attack may be extended to extract a global descent, hence $V$ can be uncovered, if the following special conditions are satisfied: $\dim L_S = g$, $\dim L_{\hat{S}} = dg$ and the linear part of the defining set of $g$ polynomials are linearly independent. The idea is by Gaussian elimination we may assume that one of the defining polynomial $F$ has the linear part with coefficient 0 in $g-1$ variables.
 Setting the descents of the $g-1$ variables to 0 leads to $d(g-1)$ linear conditions.  This implies the polynomials in $\hat{F}$ are likely in a subspace of $L_{\hat{S}}$ of dimension $dg -d(g-1) =d$.  Hence $\hat{F}$ can be extracted just like the linear case discussed before.

In general $\dim L_S \ge g$. When $\dim L_S > g$, the attack does not work even if the linear part of the $g$ defining polynomials are linearly independent.
We say that $L_S$ is {\em tight} if $\dim L_S =g$.

More generally the linear-term attack works when a support set $S'$ can be identified together with a variable $x_i\in S'$ such that
 $\dim L_{S'} = 1$, $\dim L_{S'-\{x_i\}} = 0$ and $\dim L_{\hat{S'}}=d$.  Then there is some $F\in L_{S'}$ of the form $x_i+F'$ with $F'\in L_{S'-\{x_i\}}$.  Let $\hat{x}_i = (y_{ij})_{j=1}^d$.  Then the $d$ polynomials in $\hat{F}$
are all in $L_{\hat{S'}}$. They are clearly linearly independent and can be extracted one by one by setting $d-1$ variables to 0 as discussed before.

It is interesting to consider the linear-term attack on a map $\varphi: V \to \bar{k}$ that can be defined by a polynomial $F$.  In this situation $\varphi$ is hidden in a specification of $\hat{\varphi}$ as discussed before.
Consider
the graph $V'$ of $\varphi$, that is
$V'=\{ (x,y) : y=\varphi (x), x\in V(\bar{k})\}$.
If the support of $F$ can be bounded by some $S$ then
let $S'=S\cup \{y\}$ where $y$ is a new variable.  Consider $L_{S'}$, $L_{S'-\{y\}}$ in reference to the ideal of $V'$, and $L_{\hat{S'}}$ in reference to $\hat{V'}$.

Let $D$ be the degree of the defining set of polynomials for $V$.  The analysis below shows that linear-term attack cannot apply when $\deg F$ is substantially larger than $D$.  The attack may apply when $\deg F$ is smaller than $D$.

Any polynomial $F'$ such that $F'-F$ is in the ideal of $V$ defines the same map $\varphi$ on $V$, and $y-F'$ is in the ideal of $V'$.     If $\deg F < D$ then $S$ can be chosen to be smaller than the support of the defining set.  If there is no polynomial in the ideal of $V$ with support bounded by $S$, then $F$ is the only polynomial of support bounded by $S$ that can define the map $\varphi$ on $V$.  In this case $\dim L_{S'} = 1$, $\dim L_{S'-\{x_i\}} = 0$, so if in addition $\dim L_{\hat{S'}}=d$ then
linear-term attack applies.

If $\deg F \ge D$ and $S$ contains the support of the defining set of polynomials for $V$, then there are other polynomials $F'$ supported by $S$ such that $y-F' \in L_{S'}$, hence $\dim L_{S'} >1$.  In this case linear-term attack cannot apply.

We now describe an attack which shows that polynomial number of sampled points on the descent variety may contain enough information for us to determine the descent table.  However the attack is practical only when the degree $d$ of extension of $K$ over $k$ is constant.

If $F\in R$ is supported by $S$ and for all $\alpha$ such that $\hat{\alpha}$ is a sampled point, $\hat{F} (\hat{\alpha})=0$.  Then the $d$ polynomials in $\hat{F}$ are all in $L_{\hat{S}}$.
To find a $\hat{F}\in L_{\hat{S}}$, write $F=\sum_{i=1}^t a_i m_i$ with $a_i\in K$ treated as unknown and $m_i \in S$.  Then we can express each polynomial in $\hat{F}$ in terms of the unknown ${\gamma_{ijk}}$ in the descent table and the $dt$ unknown
$a_{ij}$ with $\hat{a}_i = (a_{ij})_{j=1}^d$ for $i=1,\ldots,t$.

For all $\alpha$ such that $\hat{\alpha}$ is a sampled point, $\hat{F} (\hat{\alpha})=0$.     Each point $\hat{\alpha}$ gives us $d$ polynomial conditions, if we have $N$ sampled points where $dN > d^3 + dt$ we may have enough conditions to define a zero-dimensional polynomial system, solving which gives us a finite number of possible choices for the descent table.  However this attack is not practical in our situation where $d$ is large.

\section{A concrete construction}
\label{jac}
We apply the general idea to a more concrete setting where we take $A$ to be the jacobian variety of a hyperelliptic curve $C$ of genus $g$ with an affine model $y^2 = f(x)$ where $f\in K[x]$ of degree $2g+1$ where $g >1$.  Again let $d = [K:k]$.

We do not consider the case $g=1$ since in this case $A$ has an affine model defined by a cubic polynomial with a linear term, hence a relatively weak case in light of the analysis in \S~\ref{analysis}.

We follow \cite{C} and consider the birational model for representing points of $A$ by reduced divisors on $C$.  Following \cite{C}, a {\em semireduced} divisor is of the form $\sum_{i=1}^r P_i -r \infty$, where if $P_i = (x_i,y_i)$ then $P_j \neq (x_i,-y_i)$ for $j\neq i$.  A semireduced divisor $D$ can be uniquely represented by a pair of polynomials $(a,b)$
such that $a(x)=\prod_{i=1}^r (x - x_i)$, $\deg (b) < \deg (a)$ , and $b^2 \equiv f \mod a$. We write $D=\Div (a,b)$. The divisor $D$ is $K$-rational if  $a,b\in K[x]$.  A reduced divisor is a semireduced divisor $D$ with $r\le g$, represented by a pair of polynomials $(a,b)$ where $\deg b < \deg a \le g$ and $a$ is monic.  If $D$ is $K$-rational then $a,b\in K[x]$, and $(a,b)$ can be naturally identified with a point in $K^{2g}$.

The addition law can be described in terms of two algorithms: {\em composition} of semireduced divisors and {\em reduction} of a semireduced divisor to a reduced divisor \cite{C}.

Suppose $D_1 =\Div (a_1, b_1)$ and $D_2 = \Div (a_2, b_2)$ are two semireduced divisors.  Then $D_1 + D_2 = D + ( h )$ where $D = \Div (a,b)$ is semireduced and $h(x)$ is a function, and $a,b$ and $h$ can be computed by a
{\em composition} algorithm.  We have
\[ h = gcd (a_1, a_2,b_1+b_2 ) = h_1 a_1 + h_2 a_2 + h_3 (b_1+b_2)\]
where $h_1$, $h_2$ and $h_3$ are polynomials.

\[ a = \frac{a_1 a_2}{h^2} \]

\[ b = \frac{h_1 a_1 b_2 + h_2 a_2 b_1 + h_3 (b_1 b_2 +f)}{h} \mod a \]

Suppose $D=\Div (a,b)$ is a semireduced divisor with $\deg a > g$.  Then
$D + (y-b) = E=\Div (a',b')$ where $\deg a' \le \deg a -2$ and $E$ is semireduced.  We have
\[ a' = \frac{f-b^2}{a}\]
\[b'= -b \mod a' .\]

If $D_1$ and $D_2$ are two reduced divisors then after a composition we get a semireduced divisor of degree at most $2g$.
So in $O(g)$ iterations of reductions we eventually obtained a reduced divisor $D_3$ and a function $h$ so that
$D_1 + D_2 = D_3 + (h)$.   We call this computation {\em addition}:  on input reduced divisors $D_1=\Div (a_1,b_1)$ and $D_2 = \Div (a_2,b_2)$, a reduced divisor $D_3 = \Div (a_3,b_3)$ together with a function $h$ are constructed, so that $D_1 + D_2 = D_3 + (h)$.

Note that the function $h$ is of the form $\frac{h_1}{h_2}$ where $h_1 (x)$ is a polynomial of degree less than $2g$ and
$h_2$ is the product of $O(g)$ functions of the form $y-\beta(x)$ where the degree of $\beta (x)$ is less than $2g$.
We observe that the basic operations in composition and reduction are polynomial addition, multiplication and division (to obtain quotient and remainder).  Addition and multiplication are linear and quadratic in the coefficients of the input polynomials respectively.  Consider polynomial division.  Let $f$ and $g$ be polynomials of degrees $n$ and $m$ respectively.   Then $f= qg+r$ where $\deg q = n-m$ and $\deg r \le m-1$.  Let $(f_i)_{i=0}^n$, $(g_i)_{i=0}^m$, $(q_i)_{i=0}^{n-m}$ and $(r_i)_{i=0}^{m-1}$ be the coefficient vectors of $f,g,q,r$ respectively.  Assume without loss of generality $g$ is monic so that $g_m=1$.  Then $q_{n-m-i}$ can be expressed as a polynomial in $f_i$'s and $g_i$'s of degree $i+1$, for $i=0,\ldots, n-m$; and $r_i$ can be expressed as a polynomial of degree $n-m+2$ for $i=0,\ldots, m-1$.

A point on the jacobian of $C$ is represented by a reduced divisor $\Div (a,b)$ where $a$ is monic, $\deg a \le g$ and $\deg b < \deg a$, satisfying $f\equiv b^2 \mod a$.  The last condition can be expressed by demanding the remainder of the division of $f-b^2$ by $a$ to be 0.  From the discussion above this translates into $\deg a$ polynomial conditions of degree $O(g)$, namely by setting the $\deg a$ many remainder polynomials to zero.  We have $O(g^2)$ affine pieces depending on $\deg a $ and $\deg  b$.   It can be shown that for most cases of $f$, $L_S$ is not tight where $S$ is the support of the remainder polynomials.  Hence the linear-term attack does not apply in our setting as we consider the descent variety of $A$.

The addition of two reduced divisors involves $O(g)$ polynomial divisions.  Each division leads to $O(g)$ branches of computation depending on the degree of the remainder.  The degrees of the coefficients of quotient and remainder polynomials as polynomials in the coefficients of $a_1$, $b_1$, $a_2$ and $b_2$ increase by a factor of $O(g)$ with each division.  A routine analysis shows that the addition of reduced divisors can be divided into
$g^{O(g)}$ cases.  Each case is a morphism defined by $O(g)$ polynomials of degree $g^{O(g)}$ on an algebraic set, and the algebraic set is defined by $g^{O(1)}$ polynomials of degree $g^{O(g)}$.  In each case the function $h$ is the product of $O(g)$ functions, each with coefficients expressed as polynomials of degrees $g^{O(g)}$ in the coefficients of $a_1$, $b_1$, $a_2$ and $b_2$.  More precisely, as mentioned before, $h$ is of the form $\frac{h_1}{h_2}$ where $h_1 (x)$ is a polynomial of degree less than $2g$ and $h_2$ is the product of $O(g)$ functions of the form $y-\beta(x)$ where the degree of $\beta (x)$ is less than $2g$.  In this form it is suitable for evaluation at points but not reduced divisors of the form $\Div (a,b)$, which is needed for pairing computation.  Hence more work on $h$ is needed.

From the discussion above we know that the component functions $h_i$ in constructing $f$ is built from polynomials in $x$ of degree less than $2g$ and polynomials of the form  $y-b(x)$ where $\deg b < 2g$.  We need to process these polynomials so that we can evaluate $h$ at reduced divisors in the computation of the pairing $e$.

Let $\nuinf$ denote the valuation on the function field of $C$ at infinity. Then $\nuinf(x)=-2$ and $\nuinf(y)=-(2g+1)$,
and $x^g y^{-1}$ is a local uniformizing parameter for $\nuinf$.

For functions $f$ and $g$ we write $f\siminf g$ if $\frac{f}{g} (\infty)=1$.

For $f\in K[x]$ let $f_{\infty}$ denote the leading coefficient of $f$.  Then $\nuinf (f) = -2\deg f$ and
$f \siminf f_{\infty} x^{\deg f}$.

Consider the function $y-b$ where $b\in K[x]$.  If $\deg b \le g$ then
$\nuinf (y-b) = \nuinf (y)= -(2g+1)$, and $\nuinf (y^{-1} b) > 0$.  We have
$\frac{y-b}{y} (\infty) = (1 - y^{-1} b ) (\infty) = 1$, so $y\siminf y-b$.

If $\deg b > g$ then $\nuinf (b^{-1} y ) > 0$.  We have $\frac{y-b}{b} (\infty) = (b^{-1} y -1 )(\infty) =-1$, so
$y-b\siminf -b$.

Recall in adding two reduced divisors $D_1=\Div(a_1,b_1)$ and $D_2=\Div(a_2,b_2)$, we have $D_1 + D_2 = (h) + D_3$ with $D_3$ reduced and $h$ is of the form
$\frac{h_1}{h_2}$ where $h_1 \in K[x]$ is of degree less than $2g$ and
$h_2 = \prod_i y - \beta_i (x) $ where $\deg \beta_i$ and the number of $i$ are both less than $2g$.
Let
\[ h_{\infty} = \frac{(h_1)_{\infty}}{ \prod_{i, \deg \beta_i > g} -(\beta_i)_{\infty} } .\]
Then
\[ h\siminf h_{\infty} y^{-a} x^c\]
where $a$ is the number of $i$ such that $\deg \beta_i \le g$ and $c=\deg h_1 -\sum_{i,\deg \beta_i > g} \deg \beta_i$.
Recall from earlier discussion that $(h_1)_{\infty}$ and $(\beta_i)_{\infty}$ can be expressed as polynomials in the coefficients of $a_1$, $a_2$,  $b_1$ and $b_2$ of degree $g^{O(g)}$.

Consider now the evaluation of $h$ at the affine part of a reduced divisor.

Let
\[ D=\Div (a',b') = \sum_i P_i - r\infty\] be a reduced divisor.
Then $y(P_i) = b' (P_i)$, so
\[ (y-b) (\sum_i P_i ) = (b' -b) (\sum_i P_i ) =\prod_i (b'-b) (\alpha_i)\]
where $a' (x) = \prod_i (x-\alpha_i)$.

Let $\Phi(x) =\sum_{i=0}^{2g-1} u_i x^i\in A[x]$ where $A = K[ u_0,\ldots,u_{2g-1}]$ and the $u_i$ are variables.
We can construct by the fundamental theorem of symmetric polynomials a polynomial $G (u_0,\ldots, u_{2g-1}, t_1,\ldots,t_g)$ such that
\[G(u_0,\ldots,u_{2g-1}, s_1,\ldots,s_g) = \prod_{i=1}^g \Phi(z_i)\]
where $s_i$ is the $i$-th symmetric expression in $z_1,\ldots,z_g$ ($s_1 = z_1 + \ldots+z_g$ for example).
The polynomial $G$ has degree $O(g)$ in $u_0,\ldots,u_{2g-1}$ and degree $O(g)$ in $t_1,\ldots,t_g$.

If $f=\sum_{i=0}^m a_i x^i \in K[x]$ of degree $m < 2g$.   Denote by $G_f$ the polynomial obtained by specializing $G$ at
$u_i = a_i$ for $i=0,\ldots,m$ and $u_i = 0$ for $i > m$.   Thus
\[ G_f (s_1,\ldots, s_g) = G(a_0,\ldots,a_m,0,\ldots,0,s_1,\ldots,s_g).\]
If
\[ \rho (x) = \prod_{i=1}^r (x-\gamma_i )=x^r+ \sum_{i=1}^r (-1)^i c_i x^{r-i}\] with $c_i \in K$ and $r\le g$, then
$c_i = s_i (\gamma_1,\ldots,\gamma_g)$,and
\[ \prod_{i=1}^r f(\gamma_i) = G_f (c_1,\ldots,c_r,0\ldots,0).\]

If $D=\Div (a,b)$ is a reduced divisor then $D= D^{+} - r\infty$ for some $r\le g$. Write $a (x) = x^r + \sum_{i=1}^r a_i x^{r-i}$ and $b (x)=\sum_{i=0}^{r-1} b_i x^i$.

If $f\in K[x]$ is of degree less than $2g$, then
\[ f(D^{+}) =  G_f (c_1,\ldots,c_r,0,\ldots, 0)\]
where $c_i = (-1)^i a_i$.

For function $y-\beta(x)$ where $\deg \beta <2g$, then
\[ (y-\beta) (D^{+} ) = G'_{\beta} (c_1,\ldots,c_r,0,\ldots,0, b_0,\ldots,b_r,0,\ldots,0)\]
where
$G'_{\beta} (u_1,\ldots,u_g, b_0,\ldots,b_{g-1}) = G_{b-\beta} (u_1,\ldots,u_g)$ is $G$ specialized at $b-\beta$ while treating the coefficients of $b$ as unknown.

Recall again in adding two reduced divisors $D_1=\Div(a_1,b_1)$ and $D_2=\Div(a_2,b_2)$, we have $D_1 + D_2 = (h) + D_3$ with $D_3$ reduced and $h$ is of the form
$\frac{h_1}{h_2}$ where $h_1 \in K[x]$ is of degree less than $2g$ and
$h_2 = \prod_i y - \beta_i (x) $ where $\deg \beta_i$ and the number of $i$ are both less than $2g$.
Therefore by specializing $G$ to $h_1$ and to $b-\beta_i$ and taking product we can form $A(u_1,\ldots,u_g )$ and $B(u_1,\ldots,u_g, v_0,\ldots,v_{g-1})$ of degree $O(g^2)$, and each coefficient of $A$ and $B$ is a polynomial in the coefficients of $a_1$, $a_2$, $b_1$ and $b_2$ of degree $g^{O(g)}$, such that if $D=\Div (a,b)$ is a reduced divisor and
$D=D^+ - r\infty$ with $D^+$ positive, then $h(D^+ )$ can be computed by evaluating $A$ and $B$ with $u_1$, ..., $u_g$ being the coefficients of $a$ padded with 0 if necessary, and $v_0$,...,$v_{g-1}$ the coefficients of $b$, padded with 0 if necessary.

In summary, the algebraic program for the addition computes a morphism $m: A(\bar{k})\times A(\bar{k}) \to A(\bar{k})$, a function
$G: A(\bar{k})\times A(\bar{k}) \times A(\bar{k}) \times A(\bar{k}) \to \bar{k}$, and another function
$G_{\infty} : A(\bar{k}) \times A(\bar{k}) \to \bar{k}$.
On input reduced divisors $D_1=\Div(a_1,b_1)$ and $D_2=\Div(a_2,b_2)$, if $D_1 + D_2 = (h) + D_3$ where $D_3$ is reduced.  Then $m (D_1, D_2 ) = D_3$, $G(D_1, D_2, D) = h(D^+)$ where $D^+$ is the positive part of the reduced divisor $D$, and
$G_{\infty} (D_1,D_2) = h_{\infty}$.  The program  can be divided into
$g^{O(g)}$ cases.  In each case each coefficient of $a_3, b_3$ in the resulting reduced divisor $D_3=\Div (a_3, b_3)$ can be expressed as a polynomial of degree $g^{O(g)}$ in the coefficients of $a_1$, $b_1$, $a_2$ and $b_2$.
Let $h$ be such that  $D_1 + D_2 = (h) + D_3$.  Then $h_{\infty}$, $A(u_1,\ldots,u_g )$ and $B(u_1,\ldots,u_g, v_0,\ldots,v_{g-1})$ as discussed above can be formed so that we can evaluate $h$ at reduced divisors.  The polynomials $A$ and $B$ are of degree $O(g^2)$, with each coefficient being a polynomial in the coefficients of $a_1$, $a_2$, $b_1$ and $b_2$ of degree $g^{O(g)}$, and  $h_{\infty}$
can be expressed as a fraction of two polynomials of degree $g^{O(g)}$ in the coefficients of $a_1$, $a_2$,  $b_1$ and $b_2$.

 The pairing defined by Weil reciprocity is suitable for our application. We describe its computation below.  If a reduced divisor $D$ represents an $\ell$-torsion point, then $\ell D$ is the divisor of a function $f$.  Given two reduced divisors $D_1$ and $D_2$ that represent two $\ell$-torsion points, we define the pairing to be
\[ e (D_1, D_2 ) = \frac{f_1 (D_2)}{f_2 (D_1)}\]
where $\ell D_i = (f_i)$ for $i=1,2$.

Suppose $D$ is a $\ell$-torsion reduced divisor.  We recall how to efficiently construct $f$ such that $\ell D = (f)$ through the squaring trick \cite{Miller,Miller1}.

Apply addition to double $D$, and get
\[ 2D = (h_1) + D_1 \] where $D_1$ is reduced.
Inductively, we have $H_i$ such that
\[ 2^i D = (H_i) + D_i \] with $D_i$ reduced.
Apply addition to double $D_i$ and get
\[ 2D_i = (h_{i+1}) + D_{i+1}\] with $D_{i+1}$ reduced.  Then
\[ 2^{i+1} D = (H_{i+1}) + D_{i+1}\]
where $H_{i+1} = H_i^2 h_{i+1}$.

Write $\ell = \sum_{i} a_i 2^i$ with $a_i \in \{0,1\}$.
There are $O(\log \ell)$ non-zero $a_i$.  So apply $O(\log\ell)$ many more additions and we can construct $h$ such that
$\ell D = (h)$.
From the construction of $h$ we see that $f\siminf f_{\infty} y^r x^s$ for some integers $r,s$ and $f_{\infty}$ can be calculated from $(h_i)_{\infty}$ easily.
Given a reduced divisor $D= D^+ - r\infty$, $h(D^+)$ can be evaluated efficiently using the pairs of polynomials associated with the $h_i$'s.

Given reduced $\ell$-torsion divisors $D_1$ and $D_2$, we construct $f_1$ and $f_2$ such that $(f_1)=\ell D_1$ and
$(f_2)=\ell D_2$.  Then $e(D_1,D_2)= f_1 (D_2)/ f_2 (D_1)$. Write $D_1 = D_1^{+} - r_1 \infty$ and $D_2 = D_2^{+} - r_2\infty$.  Then $\nuinf f_i = -\ell r_i$ for $i=1,2$. So $\nuinf (f_1^{-r_2} f_2^{r_1} ) =0$. We see that
\[ \frac{f_1 (-r_2\infty)} {f_2 (-r_1\infty)} = \alpha^{-r_2} \beta^{r_1} \]
where $\alpha=(f_1)_{\infty}$ and $\beta = (f_2)_{\infty}$.

As discussed above $f_1 (D_2^+)$ and $f_2 (D_1^+)$ can be evaluated efficiently.  Hence $e(D_1,D_2)$ can be computed efficiently.

We now describe how the pairing can be extended to $\hat{A}'[\ell]\times\hat{A}[\ell]$.

Let $A_i = A^{\sigma^i}$ for $i=0,\ldots,d-1$.  Then $e$ can be naturally extended to $e_{i}: A_i [\ell] \times A_i [\ell]$ through the natural map $A\to A_i: \alpha\to\alpha^{\sigma^i}$, so that
$e_{i} (D_1, D_2 ) = e (D_1^{\sigma^{-i}}, D_2^{\sigma^{-i}} )$.

Define $E: \hat{A}'[\ell]\times \hat{A}[\ell]$ such that for $D_1\in\hat{A}'[\ell]$ , $D_2\in \hat{A}[\ell]$,
\[ E(D_1,D_2) = \prod_{0\le i\le d-1} e_{i} (\delta'^{\sigma^i}(D_1), \delta^{\sigma^i} (D_2))
=\prod_{0\le i\le d-1} e(\delta'(D_1^{\sigma^{-i}}), \delta (D_2^{\sigma^{-i}})).\]

One can verify that $E$ is bilinear and skew-symmetric using the fact that $e$ is.

In order to compute $e(\delta' {D'}_1^{\sigma^{-i}} , \delta D_2^{\sigma^{-i}} )$ on input $D'_1$ and $D_2$, we need a corresponding twisted version of the descent of $G$ and $G_{\infty}$.   They are defined in the following.

For $D_1 , D_2 \in\hat{A}[\ell]$ and $D'\in\hat{A}'[\ell]$, $G^{(i)} (D_1, D_2, D' ) = G ( \delta {D}_1^{\sigma^{-i}}, \delta D_2^{\sigma^{-i}}, {\delta'} {D'}^{\sigma^{-i}} )$.

For $D_1 , D_2 \in\hat{A}[\ell]$, $G^{(i)}_{\infty} (D_1,D_2) = G_{\infty} ( \delta {D}_1^{\sigma^{-i}}, \delta D_2^{\sigma^{-i}} )$.

Similarly where $D'_1, D'_2\in\hat{A}'[\ell]$ and $D\in\hat{A}[\ell]$,
$G'^{(i)} (D'_1, D'_2, D ) = G ( \delta' {D'}_1^{\sigma^{-i}}, \delta' {D'}_2^{\sigma^{-i}}, \delta D^{\sigma^{-i}} )$.

For $D'_1,D'_2\in\hat{A}'[\ell]$,
$G'^{(i)}_{\infty} (D'_1,D'_2) = G_{\infty} ( \delta' {D'}_1^{\sigma^{-i}}, \delta' {D'}_2^{\sigma^{-i}} )$.

Let $m$ be the addition morphism on $A$ and $\hat{m}$ its descent on $\hat{A}$.
Suppose $D_1,D_2\in\hat{A}(\bar{k})$ and $D_3 = \hat{m} (D_1, D_2)$.  So
$\delta D_1^{\sigma^{-i}}  + \delta D_2^{\sigma^{-i}}    = (h_i) + \delta D_3^{\sigma^{-i}} $ on $A$ for some function $h_i$.
Then $(h_i)_{\infty} = G_{\infty} ( \delta {D}_1^{\sigma^{-i}}, \delta D_2^{\sigma^{-i}})=G^{(i)}_{\infty} (D_1,D_2)$.
For $D'\in\hat{A}'(\bar{k})$, $h_i (\delta' {D'}^{\sigma^{-i}}  )  = G ( \delta{D}_1^{\sigma^{-i}}, \delta D_2^{\sigma^{-i}}, {\delta'} {D'}^{\sigma^{-i}} )=G^{(i)} (D_1, D_2, D' )$.

Let $m$ be the addition morphism on $A$ and $\hat{m}'$ its descent on $\hat{A}'$.
Suppose $D'_1,D'_2\in\hat{A}'(\bar{k})$ and $D'_3 = \hat{m}' (D'_1, D'_2)$.  So
${\delta'} {D'}_1^{\sigma^{-i}}  + {\delta'} {D'}_2^{\sigma^{-i}}    = (h_i) + {\delta'} {D'}_3^{\sigma^{-i}} $ on $A$ for some function $h_i$.
Then $(h_i)_{\infty} = G_{\infty} ( \delta' {D'}_1^{\sigma^{-i}}, \delta' {D'}_2^{\sigma^{-i}})=G'^{(i)}_{\infty} (D'_1,D'_2)$.   For $D\in\hat{A}(\bar{k})$, $h_i ({\delta} D^{\sigma^{-i}}  )  = G ( \delta' {D'}_1^{\sigma^{-i}}, \delta' {D'}_2^{\sigma^{-i}}, {\delta} D^{\sigma^{-i}} )=G'^{(i)} (D'_1, D'_2, D )$.

From this and the discussion before we see that on input $D'_1\in\hat{A}'[\ell]$ and $D_2$ in $\hat{A}[\ell]$, $E (D'_1,D_2)$ can be computed with $O(\log \ell)$ application of $\hat{m}$, $\hat{m}'$, $G^{(i)}$,  $G^{(i)}_{\infty}$, $G'^{(i)}$,  $G'^{(i)}_{\infty}$, $i=0,\ldots,d-1$.

To construct a trilinear map, we find $\ell$-torsion reduced divisors $D_{\alpha}$ and $D_{\beta}$ on $\hat{A}$ along with
nontrivial $\lambda, \mu \in \End (\hat{A}[\ell])$ such that $\lambda (D_{\beta}) = D_{\alpha}$, and $\mu (D_{\beta})= 0$ on $\hat{A}$ and
$e(D_{\alpha}, D_{\beta}) \neq 1$.
We make sure that $D_{\alpha}$ and $D_{\beta}$ are not the descents of points on $A[\ell]$.

Specify a set $\Sigma$ of $d^{O(1)}$ maps $\varphi_i$ where $\varphi_i = \varphi_{M_i}$ for some $d\times d$, (0,1)-matrix where each row has at most 2 nonzero entries, so that $\lambda$ and $\mu$ can be specified as a linear sum over $\varphi_i$.  So,
$\lambda = a_0+\sum_i a_i \varphi_i$ and $\mu = b_0+\sum_i b_i \varphi_i$ with $a_i, b_i \in \F_{\ell}$

The two points together with $\lambda$ and $\mu$ can be constructed from points on $A[\ell]$ and the matrices $M_i$'s with the help of map $\rho$, as explained in \S~\ref{tri-weil}.

Let $D'_{\alpha}$ be the point in $\hat{A}'[\ell]$ corresponding to $D_{\alpha}$.

As in \S~\ref{tri-weil}, let $G_1$ the cyclic group generated by $D'_{\alpha}$ with group morphism determined by $\hat{m}'$.  Let $G_2$ the cyclic group generated by $D_{\beta}$ with group morphism determined by $\hat{m}$.

Let $\Lambda$ be the $\F_{\ell}$ associative non-commutative algebra generated by a set $\Sigma$ of $N_1$ variables $z_1,\ldots,z_{N_1}$.

Let $\phi$ be the morphism of algebra from $\Lambda$ to $\End(\hat{A}[\ell])$ such that $\phi(z_i)=\varphi_i$ for all $i$.  Then $\phi$ defines an action of $\Lambda$ on $\hat{A}[\ell]$.

Let $f_{\lambda} = a_0+\sum_i a_i z_i$ and $f_{\mu} = b_0+\sum_i b_i z_i$, so that $\phi (f_{\lambda}) = \lambda$ and $\phi (f_{\mu}) = \mu$.

For $n\in\Z_{\ge 0}$, let $\Lambda_n$ denote the submodule of $\Lambda$ spanned by monomials over $\Sigma$ of degree no greater than $n$.

Set a bound $N=O(d)$ and let $S=\{f_{\lambda}\}\cup \{ wf_{\mu}: w$ is a monomial over $\Sigma$ of degree less than $N\}$.

Let $U$ be the submodule of $\Lambda$ spanned by $S$.
Let $G_3 = U_1/U$ with $1+U$ as the generator.

For $z\in\F_{\ell}$, $z+U\in G_3$ is encoded by a sparse random representative  $\gamma\in z+U\subset U_1\subset\Lambda_N$.
More precisely,  we randomly select $t=d^{O(1)}$ elements $w_i\in S$ and random $a_i\in\F_{\ell}$, then compute $\gamma=z+\sum_i a_i w_i$ as an element in $\Lambda_N$.
Then $\gamma\in \Lambda_N$ is an encoding of $z+U\in G_3$.

The trilinear map $G_1\times G_2 \times G_3 \to \mu_{\ell}$ sends
$(xD'_{\alpha},yD_{\beta},z+U)$ to $\zeta^{xyz}$ where $\zeta=E(D'_{\alpha},D_{\beta})$.
Suppose $z+U$ is represented by $\gamma\in z+U$.  Then
\[ E (xD'_{\alpha},\phi(\gamma)(yD_{\beta}))=E (xD'_{\alpha},zyD_{\beta})=\zeta^{xyz}.\]

The sparsity constraint is to make sure that the map $\gamma$ can be efficiently executed,so that the trilinear map can be efficiently computed.

Fix a public basis $\theta_1,\ldots,\theta_d$ of $K/k$.

The points $D'_{\alpha}$ and $D_{\beta}$, maps $\varphi_i$, $i=1,\ldots,N_1$, and additional programs mentioned above are defined over $K$, published using the public basis $\theta_1,\ldots,\theta_d$.

Publish the descents $\hat{m}$ and $\hat{m}'$ of the addition $m$.   Each can be published in $g^{O(g)}$ affine pieces, and we make sure that the algebraic description does not contain any global descent.
Publish the twisted descent functions $G^{(i)}$, $G^{(i)}_{\infty}$, $G'^{(i)}$ and $G'^{(i)}_{\infty}$ for $i=0,\ldots, d-1$.
Again, we make sure that the algebraic descriptions do not contain any global descent.

Publish $D'_{\alpha}$ and $D_{\beta}$ as well as the specification of $\varphi_i$.  Publish $\lambda$ and $\mu$ as linear expressions in $\varphi_i$.

The discrete logarithm problem on $G_3$ as defined in the setting of the published trilinear map is a discrete logarithm problem with the descent basis as the secret trapdoor, and the trilinear map for efficient public identity testing.
From the published trilinear map can the descent basis be determined?  The security of the trilinear map depends on the hardness of this question.

By applying $\hat{m}$ and $\varphi_i$'s to $D_{\beta}$ one can generate many other points in $\hat{A}(K)[\ell]$. Can $\hat{A}$ be efficiently determined from the sampled points?  If so, can $\hat{A}$ be efficiently decomposed as the product of conjugate abelian varieties over $K$? From such decomposition one can likely determine the descent basis efficiently.  This raise the question:  from the published trilinear map can $\hat{A}$ be determined efficiently as the product of conjugate abelian varieties?

\section*{Acknowledgements}
I would like to thank the participants of the BIRS workshop: An algebraic approach to multilinear maps for cryptography (May 2018), for stimulating and helpful discussions.

\end{document}